\documentclass[osajnl,twocolumn,showpacs,superscriptaddress,10pt]{revtex4-1} %% use 11pt for Applied Optics
\usepackage{amsmath,amssymb,graphicx}
\begin{document}
\title{Observation of the Imbert-Fedorov effect via weak value amplification}
\author{G. Jayaswal}
\author{G. Mistura}
\author{M. Merano}\email{Corresponding author: michele.merano@unipd.it}
\affiliation{Dipartimento di Fisica e Astronomia G. Galilei, Universit\`{a} degli studi di Padova, via Marzolo 8, 35131 Padova, Italy}

\begin{abstract}Weak measurements have recently allowed for the observation of the spin-Hall effect of light in reflection or transmission, a spin dependent light beam shift orthogonal to the plane of incidence. We report here the observation of the Imbert-Fedorov shift via a weak value amplification scheme. The Imbert-Fedorov effect does not depend on the spin of the incident photon only, but it has richer polarization dependence. We prove that weak measurements allow for a complete experimental characterization of the polarization properties of this tiny optical effect.
\end{abstract}


\maketitle %% required
The behavior of a bounded beam of light in reflection differs from that exhibited by plane waves, the latter being ruled by geometrical optics. For finite-diameter light beams, diffractive corrections occur that shift the beam in directions parallel and perpendicular to the plane of incidence. The parallel shift is known as the Goos-H$\rm\ddot{a}$nchen (GH)\cite{Goos47}  shift and the perpendicular one is the Imbert-Fedorov (IF) shift \cite{Fedorov55, Imbert72}. These deviations from geometrical optics predictions can be either spatial or angular \cite{Merano09}. These effects have been extensively studied not only for total internal reflection (TIR), which is the context wherein the GH and IF effects were originally addressed, but also in partial dielectric reflection and transmission.

Among the various experimental techniques for the observation of optical beam shifts, weak measurements have proven to be very successful. This approach was used for the first time by Hosten and Kwiat \cite{Kwiat08} for the observation of the spin-Hall effect of light (SHEL) \cite{Onoda04, Bliokh06}. This phenomenon is a spin-dependent displacement, perpendicular to the plane of incidence, for photons transmitted through an interface. It can be seen as a photonic version of the spin-Hall effect in electronic systems. Weak measurements allowed for others results like the observation of SHEL in optical reflection \cite{Qin09} as well as in a plasmonic system \cite{Gorodetski12}. A spin dependent beam shift in the plane of incidence was reported as well \cite{Qin11}. Recently we observed the GH shift via a weak measurement approach \cite{Jayaswal13}. Although weak measurements were introduced in quantum mechanics, theoretical analysis proved that in the context of optical beam shifts they can be described by classical optics \cite{Aiello08, Gotte12, Dennis12, Toppel13, Gotte13}.

Here we report the weak measurement of the spatial IF effect. The effect is observed in TIR i.e. the simplest case in which the IF shift can be investigated. The IF effect does not depend only on the spin of the photon but it requires a more complete description in terms of the polarization properties of the incident light.

Our experimental set up shows a strict analogy with the original proposal of Aharonov, Albert and Vaidman (AAV) \cite{Aharonov88, Aharonov90}, which describes the weak measurement of the spin of the electron. In their paper an electron Gaussian beam is prepared in a pre-selected spin state. A tiny magnetic field gradient, acting as the weak measurement apparatus, splits the electron beam in two parts, and a Stern-Gerlach apparatus post-selects the final spin state. AAV noticed how weak measurements can be used as an amplification scheme for the observation of the tiny magnetic field gradient. We measure the weak value of the polarization of a light beam. We pre-select the polarization of a Gaussian beam that undergoes TIR. The IF shift acts as the weak measuring effect, and an analyzer post-selects the final polarization state. As a consequence of this, the displacement of the beam centroid on a position sensitive detector following the analyzer is a faithful amplification of the IF effect.
 
Even if the IF effect is completely classical, we measure it here with an experimental scheme derived from quantum mechanics. We can give a complete classical description of both the IF effect and the measurement approach, or we can alternatively describe the IF effect using a quantum mechanical description. This last approach has the merit to furnish a very good physical insight and to simplify the mathematical treatment necessary for the analysis of our experimental data. Owing to the one-to-one correspondence between the paraxial wave equation and the Schr$\rm\ddot{o}$dinger equation, the electric field  of a paraxial beam can be described in a way formally equivalent to the wave function of a nonrelativistic quantum-mechanical particle with spin 1/2. Let \textbf{\textit{e$_1$}} = (1, 0) and \textbf{\textit{e$_2$}} = (0, 1) be two unit vectors that span the transverse plane perpendicular to the beam propagation axis z (Fig.1). The polarization of a light beam can be described as a two-component spinor i.e.$\left\langle A \right|\left.\right.\equiv$\textit{a$_{1}$}\textit{\textbf{e$_{1}$}}+\textit{a$_{2}$}\textit{\textbf{e$_{2}$}} where $|a_{1}|^2+|a_{2}|^2=1$. Using the bra-ket notation $\left\langle p\left|\right.\right.$= (1, 0) is the \textit{p} polarization and $\left\langle s\left|\right.\right.$= (0, 1) is the \textit{s} polarization. If we limit ourselves (without loss of generality) to the total internal reflection case the quantum operator describing the IF shift is given by the 2x2 matrix \cite{Toppel13}
\begin{eqnarray}
IF = &-&\left(1+\cos\left(\delta\right)\right)\cot\left(\theta\right) \boldsymbol{\sigma_2}-\sin\left(\delta\right)\cot\left(\theta\right) \boldsymbol{\sigma_1}
\end{eqnarray}
where $\delta = (\delta_p -\delta_s)$, $\delta_{p}$, $\delta_{s}$ are the phase jumps of \textit{p} and \textit{s} polarized beam in TIR, $\theta$ is the angle of incidence and $\boldsymbol{\textit{$\sigma_{2}$}}$, $\boldsymbol{\textit{$\sigma_{1}$}}$ are the Pauli matrices:
\begin{equation}
\boldsymbol{\sigma_2}=
\left[\begin{array}{cc}
0 & -i \\
i & 0 \\	
\end{array}\right]
%\end{equation}
%\begin{equation}
\: \: \boldsymbol{\sigma_1}=
\left[\begin{array}{cc}
0 & 1 \\
1 & 0 \\	
\end{array}\right]
\end{equation}  
From expression (1) it is evident that the operator IF is hermitian. The component $\boldsymbol{\textit{$\sigma_{2}$}}$ is diagonal in the circular polarization basis (1/$\sqrt{2}$, $\pm$ i/$\sqrt{2}$) and the component $\boldsymbol{\textit{$\sigma_{1}$}}$ is diagonal in the (1/$\sqrt{2}$, $\pm$ 1/$\sqrt{2}$) linear polarization basis. These two components are both spatial shifts because in TIR there is no angular IF effect \cite{Aiello08}.
If we pre-select the polarization of the incident beam to be \textit{p} polarized, and we post-select the final polarization as $\left\langle\psi\left|\right.\right.$= ($\epsilon$, 1) (where $\epsilon$ is a small angle) the weak value of the IF matrix is:
\begin{equation}
\frac{\left\langle\psi\left|IF\right| p \right\rangle }{\left\langle \psi | p \right\rangle}
= -\frac{\cot\left(\theta\right)}{\epsilon}\big(\sin(\delta)+ i\left(1+\cos(\delta)\right)\big)
\end{equation}
where the real part of the weak value comes from $\boldsymbol{\textit{$\sigma_{1}$}}$  (and it represents the IF shift of a (1/$\sqrt{2}$, 1/$\sqrt{2}$) polarized beam) and the imaginary part comes from $\boldsymbol{\textit{$\sigma_{2}$}}$ (and it represents the IF shift of a ((1/$\sqrt{2}$, i/$\sqrt{2}$)) polarized beam).
If we pre-select the polarization of incident beam to be \textit{p} polarized, and we post-select the final polarization as $\left.\left\langle\phi\right.\right|$ = ($\epsilon$, i) the weak value of the IF matrix is:
\begin{equation}
\frac{\left\langle\phi\left|IF\right| p \right\rangle }{\left\langle \phi | p \right\rangle}
=\frac{\cot\left(\theta\right)}{\epsilon}\big(1+\cos(\delta)- i\sin(\delta)\big)
\end{equation}
where the real part of the weak value comes from $\boldsymbol{\textit{$\sigma_{2}$}}$ and the imaginary part comes from $\boldsymbol{\textit{$\sigma_{1}$}}$. Our experimental set up allows for measuring the two contribution ($\boldsymbol{\textit{$\sigma_{2}$}}$ and $\boldsymbol{\textit{$\sigma_{1}$}}$) of the IF shift by observing the imaginary part of these weak values. Equations (3) and (4) are in agreement with equation (58b) of ref. \cite{Toppel13}.
\begin{figure}[h]
\includegraphics[scale=0.17]{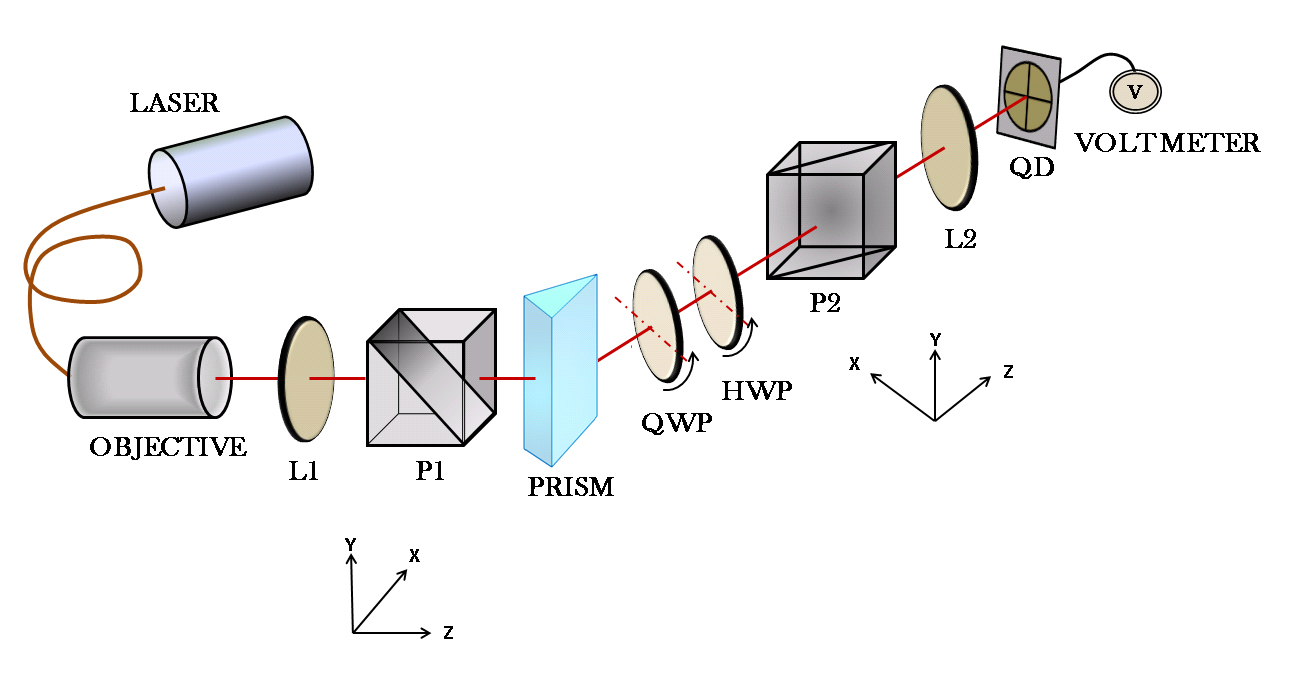}
\caption{Set up for the weak measurement of the Imbert-Fedorov shift. L1, L2: lenses with focal length 20 and 30 cm respectively. P1 and P2: Glan-Thompson polarizers. QWP: Quarter wave plate, HWP: half wave plate, QD: quadrant detector. P1 pre-selects the polarization state and the polarization analyzer composed of QWP, HWP, P2 post-selects the final polarization state.}
\end{figure}
As explained in ref. \cite{Aiello08} (just below formula (8)) these dimensionless weak values are converted into beam shifts according to the following rule (where $\left\langle y\right\rangle_{\psi}$ and $\left\langle y\right\rangle_{\phi}$ are defined as the centroids position of the post-selected  beams in the direction orthogonal to the plane of incidence): 
\begin{equation}
\left\langle y\right\rangle_{\psi} = -\frac{\lambda}{2\pi}\frac{\cot\left(\theta\right)}{\epsilon}\big(\sin(\delta)+ \frac{z}{z_{0}}(1+\cos(\delta)\big)
\end{equation}
\begin{equation}
\left\langle y\right\rangle_{\phi}= \frac{\lambda}{2\pi}\frac{\cot\left(\theta\right)}{\epsilon}\big((1+\cos(\delta)- \frac{z}{z_{0}}\sin(\delta)\big)
\end{equation}
where the real part of the complex weak value scales with {$\lambda$}/{2$\pi$} ({$\lambda$} is wavelength of the light), while the imaginary part increases with the propagation distance from the beam waist and scales with ({$\lambda$}/{2$\pi$})$\cdot(z/z_{0})$. In practice for a big enough $z/z_{0}$ ratio the imaginary part contribution of the weak values is much more amplified than the real one. A complete theoretical classical description of weak measurements of optical beam shifts can be found in ref \cite{ Aiello08} where the electric field of the reflected beam, up to the first order beyond the paraxial approximation, is computed.

In our setup the IF effect takes place when the incident Gaussian beam is totally reflected by a $45\,^{\circ}$-$90\,^{\circ}$-$45\,^{\circ}$ BK7 prism. A lens (L1) focuses the light from a collimated single-mode fiber-coupled 826 nm laser diode to a 1/$e^{2}$ intensity spot size $w_{0}$ = 60$\mu$m. The beam is \textit{p} polarized by means of a Glan-Thompson polarizing prism P1. After reflection an analyzer composed of a quarter wave plate (QWP), a half wave plate (HWP) and a Polarizer P2 is used to post-selects the desired polarization states. A lens L2  (focal length $f$ = 30 cm) collimates the beam emerging from the analyzer and directed to a quadrant detector (QD). The output signal from the QD is read by a voltmeter. In our set up the propagation distance $z=f=22\cdot z_{0}$. With such a $z/z_{0}$ ratio, the beam propagation distance is big enough to disregard in Eqs. (5) and (6) the addends that do not contain it, being them smaller than our experimental error.    
\begin{figure}[h]
\includegraphics[scale=0.30]{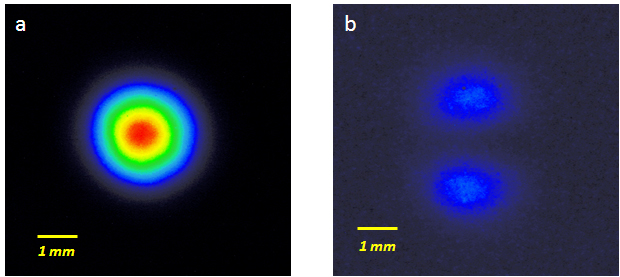}
\caption{Beam profile of the light beam transmitted through the polarization analyzer. a)  The beam profile for the weak measurements settings is still Gaussian. b) Beam profile for minimum analyzer transmission.}
\end{figure}  
The polarizer P2 is exactly orthogonal to the polarizer P1 so that it transmits \textit{s} polarized light. We first set  the QWP and the HWP in order to minimize the light transmitted through P2. In this condition the transmitted beam is not any more a Gaussian beam but has a double peak intensity profile (Fig 2b) along the y direction \cite{Sudarshan89, Ritchie91, Bliokh06}. We observe it by replacing the QD with a CCD camera. The two peaks are separated by a distance $\sqrt{2}\cdot w$ where \textit{w} is the beam waist expected for the input Gaussian beam on the CCD.

Our measurement procedure runs as follow. We perform two different measurements. In the first one we rotate the HWP of an angle (+$\epsilon$/2) (Fig. 1 shows the positive rotation direction) and then of an angle (-$\epsilon$/2), and we measure the relative position of the beam centroid in one case with respect to the other. With these settings,  the Jones matrices describing the action of the QWP ($Q_1$) and of the HWP ($H$) are:
\begin{equation}
Q_1=
\left[
\begin{array}{cc}
i & 0 \\
0 & 1 \\	
\end{array}\right]
\: \: \:\: H=
\left[
\begin{array}{cc}
1 & \pm\epsilon\\
\pm\epsilon & -1 \\	
\end{array}\right]
\end{equation}
where the $\pm$ $\epsilon$ corresponds to $\pm$ $\epsilon$/2 rotation and where we have used the approximation sin($\epsilon$)$   \: \approx \epsilon$ that it is a very good one in our case.  
This measurement procedure corresponds to post select the two final states $\left\langle s \left|HQ_1={(\pm\epsilon, i)}\right.\right.$.
With this approach we amplify the spatial IF shifts for the $\pm$$45\,^{\circ}$ linear polarized Gaussian beam. These two shifts have equal magnitude and opposite sign. Figure 3 shows our experimental data (full dots). In the set of measurements reported $\epsilon$ = 0.049 rad. The total amplification factor is (1/$\epsilon$)($z/z_{0}$) = 445. 
\begin{figure}[h]
\includegraphics[scale=0.25]{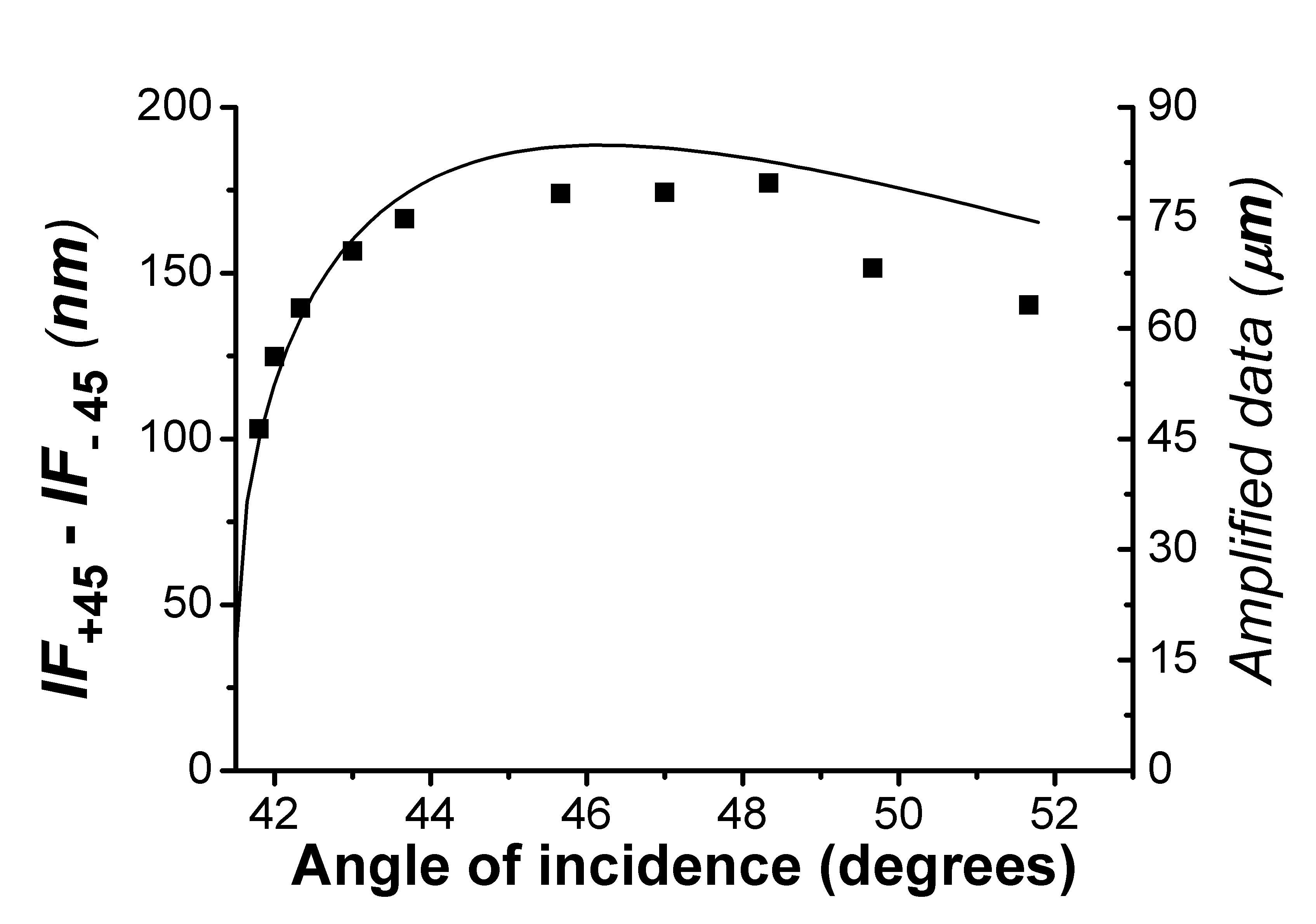}
\caption{Spatial IF beam shift for a $45\,^{\circ}$ linearly polarized incident Gaussian beam. Shown is the observed polarization-differential shift (dots), the solid line represents the theoretical prediction.The y axis on the right represents the amplified data with an amplification factor of 445.}
\end{figure}
The experimental data (dots) reported in the graph are our measurements divided by this amplification factor. The line in the graph correspond to the spatial IF shifts of a $+45\,^{\circ}$ linearly polarized beam with respect to a $-45\,^{\circ}$ linearly polarized beam in TIR. The agreement in between theory and experiment is excellent. This is the most important experimental result of this papers because it proves that weak measurements allows to observe also the linearly polarized dependent part of the IF shift. 

The second measurement scheme runs as follow: we rotate both the HWP and the QWP of an of an angle $\epsilon$/2 and an angle $\epsilon$ respectively, and we measure the relative position of the beam centroid in this case with respect to the case where both HWP and QWP are rotated respectively of -$\epsilon$/2 and -$\epsilon$.
In this case the Jones matrix for the QWP is: 
\begin{equation}
Q_2=
\left[
\begin{array}{cc}
i & \pm\left(i-1\right)\epsilon \\
\pm\left(i-1\right)\epsilon & 1 \\	
\end{array}\right]
\end{equation}
while the Jones matrix for the HWP is that given by equation (7).
This measurement procedure corresponds to post select the final states $\left\langle s \left|HQ_2={(\mp\epsilon, 1)}\right.\right.$.
\begin{figure}[h]
\includegraphics[scale=0.25]{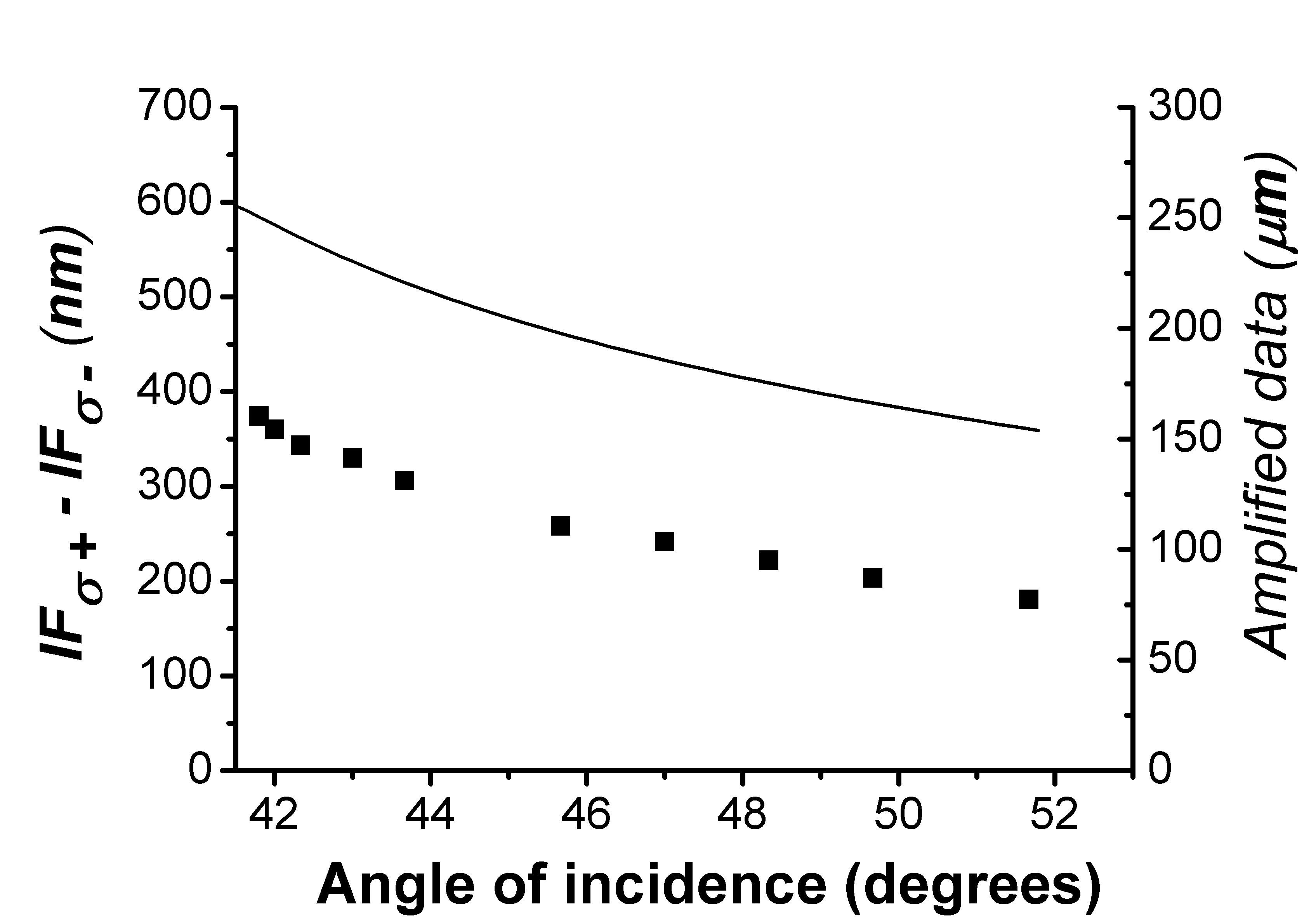}
\caption{Spatial IF beam shift for a $\pm$ circularly polarized incident Gaussian beam. Shown is the observed polarization-differential shift (dots), the solid line represents the theoretical prediction.The y axis on the right represents the amplified data with an amplification factor of 445.}
\end{figure}
With this approach we amplify the spatial IF shifts for the circular $\pm$ polarized Gaussian beam. These two shifts have equal magnitude and opposite sign. Again the amplification factor is 445, and experimental data are compared with theoretical predictions with the same procedure used for figure 3. The graph in figure 4 confirms that weak measurements provide a faithful amplification of the IF effect because experimental data and theoretical predictions have the same trend. We observe a discrepancy in the exact magnitude between data and theoretical predictions that it is still acceptable. The theory that we use for fitting does not contain free parameters and observations of optical beam shifts are delicate experiments.

\begin{figure}[h]
\includegraphics[scale=0.25]{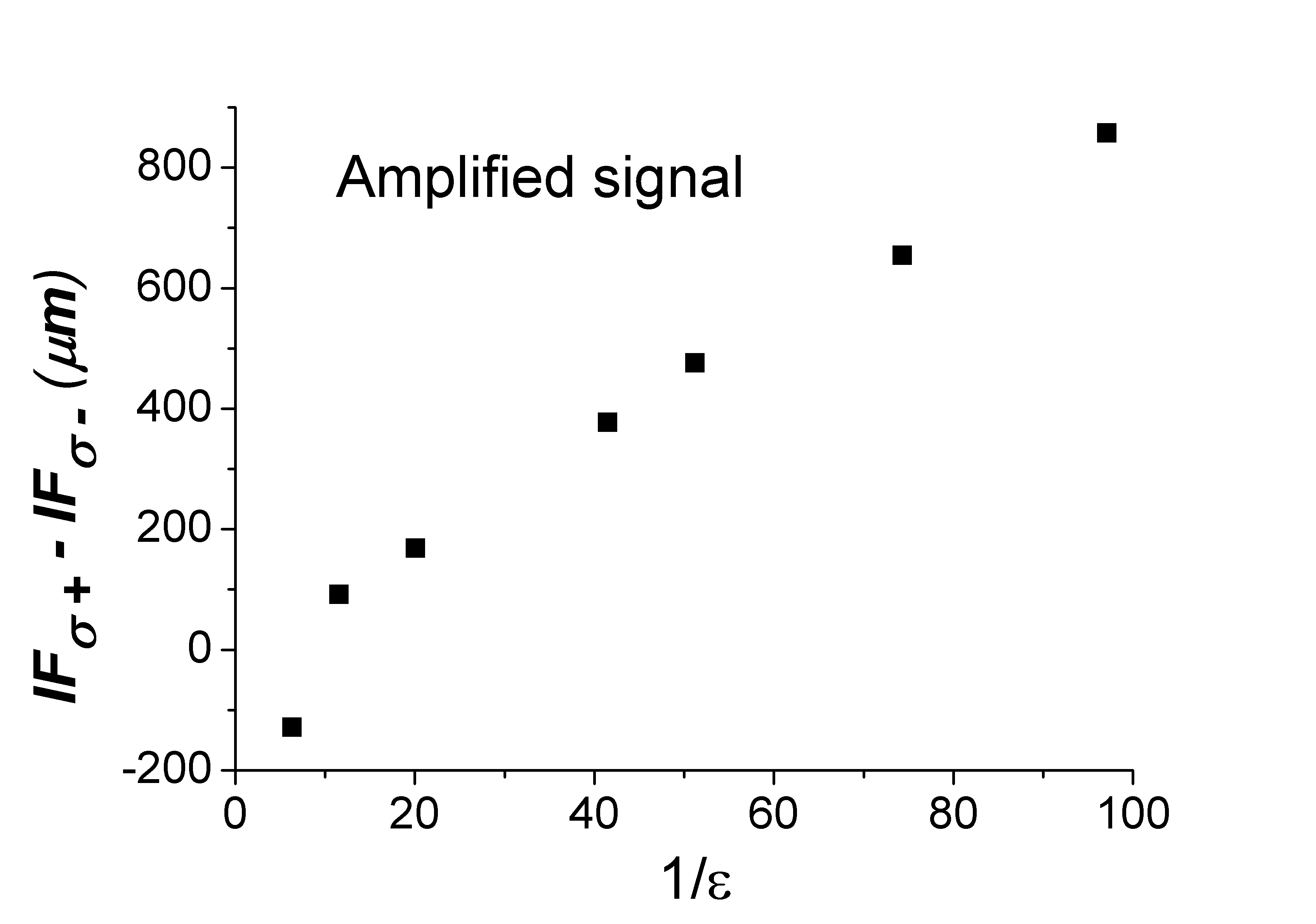}
\caption{Experimental study of the weak value amplification scheme as a function of $\epsilon$.  A  faithful amplification of the optical beam shift is observed as long as the amplified signal scale linearly with 1/$\epsilon$.}
\end{figure}

The range of values of $\epsilon$ for which the weak measurement works is quite large. Even for the smallest value of $\epsilon$ that we use in our measurements the profile of the beam transmitted through P2 is still Gaussian (Figure 2a). In figure 5 we report a series of weak measurements for different $\epsilon$ taken with post selected final states ${(\mp\epsilon, 1)}$ at an angle of incidence of $42\,^{\circ}$ and at the fixed $z/z_{0}$ = 22. As expected there is a linear dependence on 1/$\epsilon$. We see a departure from linearity at 1/$\epsilon$ = 6.4 where due to a small amplification factor the noise becomes bigger than our signal. We experimentally observed that the use of the lens L2 helps to achieve a better linear dependence on 1/$\epsilon$. We wonder if the discrepancy observed in Fig. 4 can not be partially related to this. 

In conclusion we have observed the IF shift using weak measurements in close analogy with the original AAV proposal.  Even if we observe here a spatial shift, the imaginary part of the weak value is amplified by the propagation distance. We have shown that our approach allows for measuring both the IF effect that it is diagonal in the circular $\pm$ polarization basis, that the one diagonal in the $\pm45\,^{\circ}$ linear polarization basis. 

Our paper evidences the advantages of weak measurement as an amplification scheme. While measuring spatial beam shifts of tens or hundreds of nanometers is in general a challenging task, measuring amplified light beam shifts of tens or hundreds of microns is quite easy. Weak value measurement techniques such as the one described here amplify the signal but not the technical noise (thermal, electrical, vibrational, etc.) so perturbing at the nanometer scale. The experimental approach, presented here, discriminates in between two different physical contributions to the IF effect showing a remarkable sensitivity of the weak amplification scheme in post-selection. Weak measurement is finding more and more applications recently \cite{Dixon09, Loaiza13, Puentes12, Strubi13}. We think it would be very interesting to extend our results to applications which use beam shifts as a sensitive meter \cite{Hesselink06}, such as in bio-sensing, or to beam shifts of matter waves \cite{Haan10} that are intrinsically quantum.
\begin{acknowledgements}
We gratefully acknowledge the financial support from PRAT-UNIPD contract (CPDA), 118428. 
\end{acknowledgements}

%\bibliography{letter}

\begin{thebibliography}{99}
\newcommand{\enquote}[1]{``#1''}
\bibitem{Goos47}
F.~Goos and H.~$\rm H\ddot{\textbf{a}}nchen$, Ann.\ Phys. \textbf{436}, 333
 (1947).
\bibitem{Imbert72}
C.~Imbert, Phys. Rev. D \textbf{5}, 787 (1972).
\bibitem{Fedorov55}
F.~I. Fedorov, Dokl. Akad. Nauk. Nauk SSSR \textbf{105}, 465 (1955).
\bibitem{Merano09}
M.~Merano, A.~Aiello, M.~P. van Exter, and J.~P. Woerdman, Nat. Photon.
 \textbf{3}, 337 (2009).
\bibitem{Kwiat08}
O.~Hosten and P.~Kwiat, {Science} \textbf{319}, 787 (2008).
\bibitem{Onoda04}
M.~Onoda, S.~Murakami, and N.~Nagaosa, {Phys. Rev. Lett.} \textbf{93}, 083901
  (2004).
\bibitem{Bliokh06}
K.~Y. Bliokh and Y.~P. Bliokh, {Phys. Rev. Lett.} \textbf{96}, 073903 (2006).
\bibitem{Qin09}
Y.~Qin, Y.~Li, H.~He, and Q.~Gong, {Opt. Lett.} \textbf{34}, 2551 ({2009}).
\bibitem{Gorodetski12}
Y.~Gorodetski, K.~Y. Bliokh, B.~Stein, C.~Genet, N.~Shitrit, V.~Kleiner,
  E.~Hasman, and T.~W. Ebbesen, Phys. Rev. Lett. \textbf{109}, 013901 (2012).
\bibitem{Qin11}
Y.~Qin, Y.~Li, X.~Feng, Y.-F. Xiao, H.~Yang, and Q.~Gong, {Opt. Exp.}
 \textbf{19}, 9636 (2011).
\bibitem{Jayaswal13}
G.~Jayaswal, G.~Mistura, and M.~Merano, {Opt. Lett.}
\textbf {38}, 1232 (2013).
\bibitem{Gotte12}
J.~B. $\rm G\ddot{o}tte$ and M.~R. Dennis, {New J. Phys.} \textbf{14}, 073016
  (2012).
\bibitem{Aiello08}
A.~Aiello and J.~P. Woerdman, {Opt. Lett.} \textbf{33}, 1437 (2008).
\bibitem{Dennis12}
M.~R. Dennis and J.~B. $\rm G\ddot{o}tte$, {New J. Phys.} \textbf{14}, 073013 (2012).
\bibitem{Toppel13}
F.~T{\"o}ppel, M.~Ornigotti, and A.~Aiello, {New. J. Phys.} \textbf{15}, 113059 (2013).
\bibitem{Gotte13}
J.~B. $\rm G\ddot{o}tte$, and M.~R. Dennis, {Opt. Lett.} \textbf{38}, 2295 (2013).

\bibitem{Aharonov88}
Y.~Aharonov, D.~Albert, and L.~Vaidman, {Phys. Rev. Lett.} \textbf{60}, 1351
 (1988).
\bibitem{Aharonov90}
Y.~Aharonov and L.~Vaidman, {Phys. Rev. A} \textbf{41}, 11
  (1990).
\bibitem{Sudarshan89}
I.~Duck, P.~Stevenson, and E.~Sudarshan, {Phys. Rev. D} \textbf{40}, 2112
(1989).
\bibitem{Ritchie91}
N.~Ritchie, J.~Story, and R.~Hulet, {Phys. Rev. Lett.} \textbf{66}, 1107
  (1991).
\bibitem{Dixon09}
	P.B.~Dixon, D.J.~Starling, A.N.~Jordan, and J.C.~Howell, Phys. Rev. Lett. \textbf{102}, 173601 (2009).
\bibitem{Puentes12}
	G.~Puentes, N.~Hermosa, and J.P.~Torres, Phys. Rev. Lett. \textbf{109}, 040401 (2012).
\bibitem{Loaiza13}
O.S.Maga{\~n}a-Loaiza, M.~Mirhosseini, B.~Rodenburg, and R.W.~Boyd, arXiv \textbf{1312.2981} (2013).
\bibitem{Strubi13}
	G.~Str{\"u}bi, and C.~Bruder, Phys. Rev. Lett. \textbf{110}, 083605 (2013)
\bibitem{Hesselink06}
X.~Yin and L.~Hesselink, Appl.\ Phys.\ Lett. \textbf{89}, 261108 (2006).
\bibitem{Haan10}
V.-O. de~Haan, J.~Plomp, T.~M. Rekveldt, W.~H. Kraan, A.~A. van Well, R.~M.
 Dalgliesh, and S.~Langridge, {Phys. Rev. Lett.} \textbf{104}, 010401
 (2010).

\end{thebibliography}
%\bibliographystyle{osajnl}

\end{document}